%% file: main.tex
\documentclass[a4paper,10pt]{scrartcl}
\usepackage[utf8]{inputenc}

\usepackage[a4paper,margin=2.5cm]{geometry} 

\usepackage{multirow}
\usepackage{url}
\usepackage{xcolor}
\usepackage{graphicx}
\usepackage[export]{adjustbox} 
\usepackage{subcaption}
\usepackage{float}
\usepackage{bibentry}
\nobibliography*

\usepackage[english]{babel}
\usepackage[breaklinks,pdftex,pdfusetitle]{hyperref}
\usepackage{microtype}
\usepackage{comment}
\usepackage[draft,textsize=footnotesize,textwidth=45mm]{todonotes} 
\usepackage[titletoc]{appendix}    

\usepackage{amsfonts}
\usepackage{amsmath,booktabs}

\newif\ifSED 
\SEDfalse  

\newif\ifediting
\editingtrue  

\input{macro}

\title{
DESIRE:  
A Third Way for a European Exposure Notification System}
\subtitle{Leveraging the best of centralized and decentralized systems}
\author{ Claude Castelluccia\thanks{Claude Castelluccia is the contact author. The other co-authors are listed in alphabetical order.}\\ Nataliia Bielova\\ Antoine Boutet\\ Mathieu Cunche\\ Cedric Lauradoux\\ Daniel Le M\'etayer\\ Vincent Roca\\ 
\\
PRIVATICS Team\thanks{The authors would like to thank Karthikeyan Bhargavan, Bruno Blanchet, and David Pointcheval for their comments and suggestions about this paper.}, Inria, France\\
 desire-contact@inria.fr}
\date{\today\\v1.0}

\begin{document}
\maketitle


\begin{abstract}

This document presents an evolution of the $ROBERT$ protocol that decentralizes most of its operations on the mobile devices\footnote{Available at \url{https://github.com/ROBERT-proximity-tracing/documents}}. 
$DESIRE$ is based on the same architecture than $ROBERT$ but implements major privacy improvements. In particular, it introduces the concept of \emph{Private Encounter Tokens}, that are secret and cryptographically generated, to encode encounters. In the $DESIRE$ protocol, the temporary Identifiers that are broadcast on the Bluetooth interfaces are generated by the mobile devices providing more control to the users about which ones to disclose. The role of the server is merely to match $PETs$ generated by diagnosed users with the $PETs$ provided by requesting users. It stores minimal pseudonymous data. Finally, all data that are stored on the server are encrypted using keys that are stored on the mobile devices, protecting against data breach on the server. All these modifications improve the privacy of the scheme against malicious users and authority.

However, as in the first version of $ROBERT$, risk scores and notifications are still managed and controlled by the server of the health authority,  which provides high robustness, flexibility, and efficacy.
  
\end{abstract}

\newpage
\tableofcontents
\newpage


\input{dagobert3}

\bibliography{biblio}
\bibliographystyle{plain}


\begin{appendices}

\input{appendix}

\input{Bluetooth_SCAN_RSP}

\end{appendices}

\end{document}

%% file: macro.tex
\def\atrisk{{at risk of exposure}}

\def\ServerDB{\emph{IDTable}}


%% file: dagobert3.tex
%
\section{Introduction}
\subsection{DESIRE: Leveraging the best of centralized and decentralized systems}

Recent analysis have shown that current centralized and decentralized digital contact tracing proposals come with their own benefits and risks \cite{analysis,cryptoeprint:2020:531}.

Centralized systems protect against malicious users,
but a malicious server could abuse the system by potentially re-identifying and tracing users' locations. 
Conversely, decentralized systems make public the ephemeral Bluetooth identifiers of diagnosed people, which could lead to mass surveillance by any malicious individual \cite{cryptoeprint:2020:531}. Furthermore, decentralized systems introduce some fundamental trade-off between availability, privacy, and integrity \cite{troncoso:hal-01673295}.

$DESIRE$ explores a "third way", similarly to other proposals such as  \cite{cryptoeprint:2020:493,epione}, that aims at combining the best of the two worlds. 
In particular, in $DESIRE$:
\begin{itemize}
    \item mobile devices generate their own identifiers, their private encounter tokens ($PET$), and keep full control over them. These $PET$ tokens are privately generated and unlinkable. This makes the scheme less vulnerable to a malicious server compared to a scheme where the pseudo-identifiers are generated by the back-end server.
    \item the server performs the matching between the $PET$ tokens of infected and requesting users without having access to their actual identifiers. Relying on a central server for the matching task improves the robustness and resiliency of the scheme against malicious users, compared to a scheme where the matching is performed by the devices themselves.
    \item the server computes the risk score which improves the efficacy of the system as the score function can be instantly adapted by the health authority according to the evolution of the pandemics. Furthermore, this allows the server to control the number of notifications that are sent out every day and prevents uncontrollable situations where a high number of users get notified at once. Finally, this also improves the security of the scheme against malicious users since users don't get access to the proximity information but only to the resulting
    risk scores.

\end{itemize}

\subsection{Private Encounter Tokens (PETs)}

As opposed to existing proximity tracing schemes, we propose an exposure notification system that is based on \emph{Private Encounter Tokens} ($PETs$), generated from \emph{Ephemeral Bluetooth Identifiers} ($EBID$). The $EBIDs$ and $PETs$ are generated and computed locally by the mobile devices and are \emph{unlinkable}. A $PET$ token uniquely identifies an encounter between two nodes and is \emph{secret}. They can be generated from any \emph{Non-Interactive Key Exchange} protocol ($NIKE$) \cite{NIKE}, which makes them private and unlinkable from the server \cite{Diffie1976}. 

In this report, the creation of the $PET$ tokens is based on the Diffie/Hellman key exchange protocol~\cite{Diffie1976}. For simplicity, and ease of reading, we describe our scheme using the multiplicative DH notation.  
However,  for efficiency reasons, we propose to implement it using an instance of the discrete logarithm on elliptic curves (Curve25519). 
We assume that the devices share the same group structure (order $p$ and generator $g$), which could be pre-configured in the application.

\begin{figure}[htb]
    \centering
    \includegraphics[width=11cm]{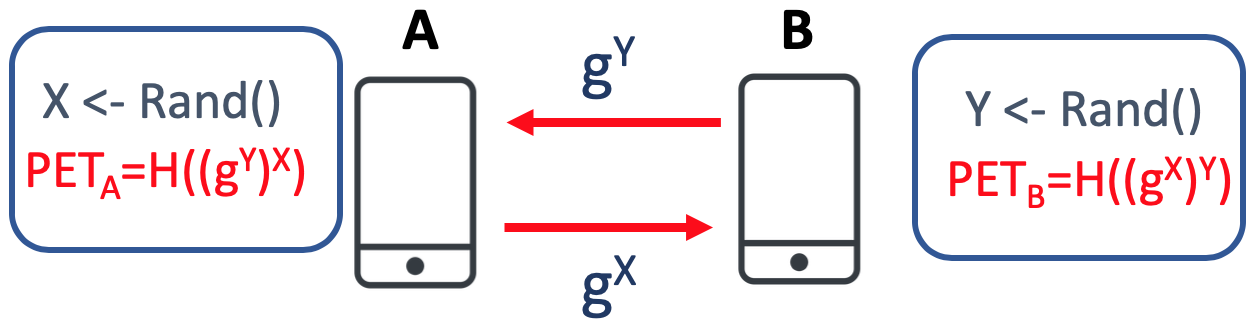}
    \caption{PET Generation}
    \label{fig:PET}
\end{figure}

At each epoch\footnote{Epochs are synchronized with the device address randomisation periods.}  $i$:

\begin{itemize}
    \item device $A$ generates and broadcasts a new $EBID$ defined as $g^{X} $, where $X$ is a secret generated by $A$.   With Elliptic Curve Cryptography and the elliptic curve $Curve25519$\footnote{\url{https://en.wikipedia.org/wiki/Curve25519}}, $g^{X}$ requires 32 bytes which are transmitted in 2 consecutive  Bluetooth packets (See Section~\ref{sec:bluetooth_com}).
    \item upon the reception of an $EBID$, $g^{Y}$, from a device $B$, device $A$ computes $(g^{Y})^X$ (which is equal to $g^{X\cdot{Y}}$) and stores in a local list\footnote{As we will see later, each node actually maintains 2 lists to prevent linkability between the PETs that are used to query the server and the PETs that are uploaded if the node is diagnosed positive.} the $PET$ token $H(g^{X\cdot{Y}})$, where $H()$ a cryptographic hash function such as SHA-256 . Similarly, device $B$ computes $(g^{X})^Y$ (which is also equal to $g^{X\cdot{Y}}$) and stores in a local list the $PET$ token $H(g^{X\cdot{Y}})$ (see Figure \ref{fig:PET}).
\end{itemize}

The $PET$ $H(g^{X\cdot{Y}})$ is a unique and a secret identifier for the encounter between $X$ and $Y$. Furthermore, the decisional Diffie–Hellman (DDH) assumption guarantees that it can only be computed by $A$ and $B$, and is therefore protected from eavesdroppers~\cite{encyclopedia}.

$PET$ has the advantage over $EBID$ that it reduces the ability of the server to link collocation information coming from different individuals. Furthermore, this method mitigates “replay attack”, where a malicious individual collects the EBIDs received by an infected (or potentially infected) individual and replays them in many locations, thus creating a large number of false positives. 

 \textbf{It is noteworthy that $PET$ tokens are not specific to $DESIRE$ and could benefit to other existing proximity-tracing proposals.}

\subsection{Protocol Overview}

All the notations used in this paper are summarized in Table~\ref{tab:notations}.
The proposed system is composed of users who install the Exposure Notification Application using $DESIRE$, $App$, and a back-end server under the control of the health authority. 
We assume that the server is configured with a well-known domain name, certificate and is highly secured. All communications with the server are performed through proxies (to hide users' network metadata that could be exploited by the server to re-identify users).



Apps interact with the system through the four following procedures:
\begin{itemize}
\item \textbf{Initialization:} 
When a user wants to use the service, she installs the application, $App$, from an official and trusted App store. $App$ then registers to the server that generates a permanent identifier (ID). \ServerDB~keeps an entry for each registered ID.  The  stored information is “de-identified" i.e., by no mean, associated to a particular identity (no personal information is stored in \ServerDB).

\item \textbf{Proximity Discovery:}
After registering to the service, each device $A$:
\begin{itemize}
    \item generates a new and unlinkable $EBID_A$ at each epoch
    \item broadcasts this $EBID_A$ regularly
    \item collects $EBIDs$ of encountered devices
    \item generates $PET$ tokens from collected $EBIDs$ if certain conditions are satisfied on, for example, contact length, received signal strength, etc.
    \item stores the generated $PET$ tokens in a local list, along with, if necessary, additional metadata (contact length, speed,...)\footnote{As described in Section\ref{sec:ed}, the device actually uses 2 lists.}.
\end{itemize}

Figure \ref{fig:proximity} shows a schematic view of the exchanges performed during the Proximity Discovery procedure described above, and will be described in greater detail in section \ref{sec:ed}.

\item \textbf{Infected User Declaration:}
    When an individual is tested 
    and {\em diagnosed} COVID-positive,
    and after an explicit user consent and authorisation (from the medical services), her smartphone's application uploads its local list of generated $PET$ tokens to the authority server, that adds them in a global list, $EList$, of exposed $PET$ tokens.

\item \textbf{Exposure Status Request:}
    $App$ queries (pull mechanism) the “exposure status" of its user by probing regularly the server with its list of generated $PET$ tokens. 
    The server then checks how many of the $App$'s tokens appear in $EList$  and computes a risk score from this information (and possibly other parameters, such the exposure duration and signal strength). 
    If this score is larger than a given threshold, the bit “1" (“\atrisk") is sent back to the $App$, otherwise the bit “0" is sent back.
    Upon reception of a message "1", a notification is displayed to the user that indicates the instructions to follow (e.g., go the hospital for a test, call a specific phone number, stay in quarantine, etc.).
\end{itemize}


\paragraph{Time-related Assumptions:}

This $DESIRE$ protocol assumes that all the smartphones and the server  are loosely time-synchronized (thanks to NTP or any other time synchronisation mechanism  like cellular mobile phone network information, or GPS time information, etc.). 
Time is expressed as the NTP “Seconds" value, which represents, for era 0, the number of seconds since 0h January 1st, 1900 UTC \footnote{\url{https://en.wikipedia.org/wiki/Network_Time_Protocol}}.

Time is discretized into epochs (e.g., of 15 minutes)\footnote{This value of 15 minutes is the rotation period of random address recommended in the Bluetooth v5.1 specification~\cite[Vol 3, Part C, App. A]{bluetooth_core_specification5.1}).}. We define as $epoch\_duration\_sec$ the duration of an epoch in seconds. Epochs are synchronized with the device address\footnote{Bluetooth \emph{device address} is sometime called \emph{MAC address}} randomisation periods.

\paragraph{Roaming Considerations:} Not considered in this version.


\begin{table}[ht]
\centering
\caption{Glossary of terms and variables used in this paper}\label{tab:notations} 
\begin{tabular}{ | p{.25\linewidth} | p{.7\linewidth}  |} 
\hline
\textbf{Name} &  \textbf{Description}   \\ 
\hline
$App$ &  Mobile application implementing $DESIRE$ \\ 
\hline
$App_A$ &  Mobile Application installed by user $U_A$ \\ 
\hline
$AT_A$ &  Authorization Token of user $U_A$ \\ 
\hline
$BLE$ &  Bluetooth Low Energy \\ 
\hline
$CT$ &  An upper-bound on the
number of days a user who has been diagnosed positive could been contagious (for example 14 days)   \\ 
\hline
$EBID$ & Ephemeral Bluetooth IDentifier  \\ 
\hline
$EBID_{A,i}$ & Ephemeral Bluetooth IDentifier generated by user $U_A$  at epoch $i$  \\ 
\hline
$EK_A$ & Encryption Key used of user $U_A$ to protect her information in table $IDTable$ \\ 
\hline
$EList$ & Exposed PET tokens list on the backend server, gathering encounters from all users diagnosed COVID-positive who uploaded their PETs \\ 
\hline
$epoch\_duration\_sec$  &  Duration of an epoch in seconds \\ 
\hline

$ERS_A$ & Exposure Risk Score of user $U_A$ \\ 
\hline
$ESR\_REQ$ & Request sent by the $App$ to query the user Exposure Status \\ 
\hline
$ESR\_REP$ & Reply message sent by the server to users to notify their Exposure Status \\ 
\hline
$ETL_A$ & Exposure Token List maintained by $App_A$ and used to upload her PETs if user $U_A$ is diagnosed COVID-positive   \\ 

\hline
$ID_A$ & Permanent and anonymous identifier of user $U_A$, stored by the server  \\ 
\hline
$IDTable$ & Database maintained by the back-end server  \\ 
\hline
$LEPM_A$ & List of Exposed PET Metadata of user $U_A$, stored in $IDTable$  \\ 
\hline
$PET$ & Private Encounter Token  \\ 
\hline

$RTL_A$ & Request Token List maintained by $App_A$ and used to upload her PETs when user $U_A$ queries her exposure status    \\ 
\hline

\hline
$SRE_A$ & Variable that indicates the last epoch when $U_A$ has sent a ”Status Request” to the server, stored in $IDTable$  \\ 
\hline
$ESR\_min$ & The minimum number of epochs between 2 consecutive $ESR\_REQ$ \\ 
\hline
$UN_A$ & Flag  indicating  if  $U_A$  has  already  been notified to be at risk of exposure. $UN_A$ is stored in the $IDTable$ \\
\hline

\end{tabular}

\end{table}

\subsection{Risk Scoring and Notification Considerations}
\label{sec:riskScoreConsiderations}
Specific and effective risk scoring is out of scope of this paper. In this document, we assume that (1) the server returns a binary reply informing users that they are at risk or not and (2) the reply is only based on a calculated risk score value. These two assumptions need to be discussed. It might be useful, for several reasons, to return a probability value instead of a binary information. Furthermore, adding some randomness in the query reply mechanism
could improve privacy (see Section \ref{sec:proba}).

We further assume that the risk assessment algorithm and its parameters should be defined and monitored by health authorities, in collaboration with epidemiologists. In addition, this algorithm should be published with all relevant information to enhance trust and understanding by the users. 
It is also of the utmost importance that the authority is able to adapt the algorithm and its parameters over time. These adjustments are necessary to take into account the evolution of the situation, in particular the number of exposed people, the number of infected people, the available resources, and also the progress made by epidemiology research to understand the virus and its spread\footnote{The need to adjust the algorithm to reflect policy changes is also stressed in \cite{NHSX}: ``It is to be expected that the optimal solution will likely involve a number of successive scenarios to reflect an early need to capture as many infections as possible and a later need to avoid quarantining of too many people as the epidemic declines and re-introductions are monitored.’’ }. Indeed, a key success factor for proximity tracing applications is their smooth integration within  the existing health care infrastructure, in particular the possibility to adapt the response to the local epidemiological situation and the available resources \cite{ECDC}. This calls for a system where the
exposure risk scores and the notifications are managed by the server. The main driver of
this approach is that it lets the health services run analytics on data and send
warnings only to those who are most at risk of having got infected. 
A major issue is the subsequent management of the alerts, within the existing proximity tracing framework (which also involves manual contact tracing, testing, quarantining, etc.). If the infrastructure is not sufficiently sized and equipped to deal with all warnings, the proximity tracing application could be useless, anxiety-provoking (because of the lack of care for individuals) and economically devastating (through the immediate, though unnecessary, withdrawal of essential workers) \cite{Pellegrini}.
The ``centralized'' approach to risk assessment and notifications makes it possible to avoid excessive warnings that could lead either to panic reactions or loss of interest or trust from the users. 

Another benefit of the central role of the server is that it increases the resilience of the system against attackers trying to identify infected users~\cite{analysis}.
To summarize, digital proximity tracing tools should be deployed as one element of a public health policy and should work in synergy with other existing measures under the
control of the public health authority. This is an essential functional requirement
of any practical exposure notification solution that seem to be incompatible with a true decentralized approach\footnote{Just to take an example, in a decentralized approach, where the risk calculation algorithm is computed on users’ devices, the health authority is not aware of the number of exposed people. This information is essential both for statistical purposes and to easily adjust the risk calculation algorithm.}.

\section{DESIRE Protocol Description}
\label{sec:protocol}

%
\subsection{Application Initialization}
\label{sec:init}

A user $U_A$ who wants to install the application on his device must download it from the official Apple or Google Play Stores. After installing the application $App_A$, $U_A$ needs to register to the back-end server.

The registration phase is composed of two phases: the \emph{authorization token generation}, during which the user obtains an anonymous authorization token, and the \emph{user registration} where the user registers to the server anonymously. 

As we will see these phases are unlinkable and provide full anonymity to user. 
\begin{enumerate}
    \item \emph{Authorization Token Generation}: In this step, the user obtains an anonymous authorization token that she uses in the User Registration phase. This could be performed by using blind signatures as described in Appendix \ref{sec:authorizationtoken}.

    \item \emph{User Registration}: Once the user, $U_A$, obtained an authorization token, $AT_A$,  she can use it to register to the server. This is performed as follows:
    \begin{enumerate}
        \item $U_A$ sends a registration message which includes his authorization token, $AT_A$.
        \item The server verifies the authorization token, creates a unique identifier $ID_A$ and an entry in its $IDTable$. The entry table contains for each registered user, the following information\footnote{This table could be extended to include additional information useful to improve the risk scoring function or necessary for the epidemiologists.}:
        
        \begin{description}

\item [-$\textrm{ID}_{A}$ (“Permanent IDentifier for A"):]
    an identifier that is unique for each registered App, and generated randomly (random draw process without replacement to avoid collision). 

\item [-$\textrm{UN}_A$ (“User A Notified"):]
    this flag indicates if the associated user has already been notified to be \atrisk\ (“true") or not (“false"). It is initialized with value “false".
    
\item [-$\textrm{SRE}_A$ (“Status Request Epoch"):]
    an integer that indicates the last epoch when $U_A$ has sent a “Status Request".

\item [-$\textrm{LEPM}_A$ (“List of Exposed PET Metadata")]:
    This list reflects the exposure of the user (temporal and frequency information) and is used to store the metadata (e.g. the duration and proximity of a contact to an infected person) necessary to compute the risk score. In this version of the protocol, each element encodes the day when the user had a contact with an infected user and the duration of that encounter. The information contained in this list can be extended to include other types of data useful to compute the risk score, such information about the environment (indoor or outdoor), signal strength,...

\item [-$\textrm{ERS}_A$ (“Exposure Risk Score")]:
    Current $U_A$'s exposure risk score. 
\end{description}

\item The server:
\begin{enumerate}
    \item generates an encryption key $EK_A$
    \item sends $ID_A$, $EK_A$ to $U_A$ 
    \item uses $EK_A$ to encrypt\footnote{We recommend the use of an authenticated encryption scheme.} all elements of $IDTable[ID_A]$, except for $ID_A$
    \item deletes $EK_A$
\end{enumerate}

\end{enumerate}

Note that the server also stores all the $AT$ tokens it has received to verify that they are only used once\footnote{In practice the server can generate authorization tokens that are only valid for a certain number of days (by changing his Public/Private key pair regularly), which would reduce the storage load on the server.}.

\end{enumerate}

%
\subsection{Encounter discovery}
\label{sec:ed}

\begin{figure}[htb]
    \centering
    \includegraphics[width=11cm]{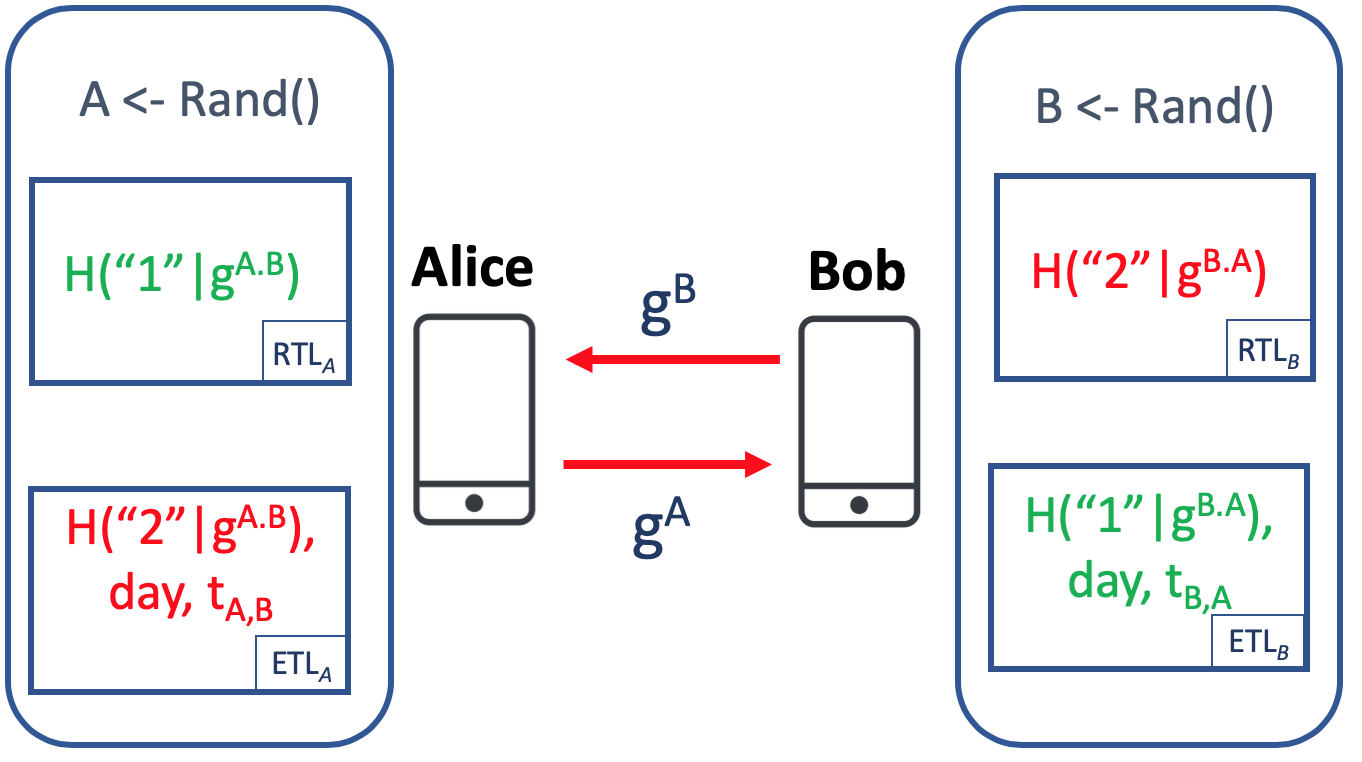}
    \caption{Device $A$ broadcasts $EBID=g^A$. Device $B$ broadcasts $EBID=g^B$. Device $A$ computes $PET^1=H("1"|g^{B.A})$ and $PET^2=H("2"|g^{B.A})$, stores $PET^1$ in $RTL_A$ and $PET^2$ in $ETL_A$. Device $B$ computes $PET^1=H("1"|g^{A.B})$ and $PET^2=H("2"|g^{A.B})$, stores $PET^1$ in $ETL_B$ and $PET^2$ in $RTL_B$.}
    \label{fig:proximity}
\end{figure}

\subsubsection{Protocol description}

Each node $A$ maintains two tables: an \emph{Exposure Token List}, $ETL_A$, and a \emph{Request Token List}, $RTL_A$.

\textbf{At each epoch $i$ of day $day$, node $A$}: 

\begin{enumerate}
    \item deletes $A_{i-1}$ and $EBID_{A,i-1}$;
    \item chooses randomly $A_i$ and computes $EBID_{A,i}=g^{A_i}$;
    \item broadcasts regularly $EBID_{A,i}$ over bluetooth; 
    \item cleans $RTL_A$ and $ETL_A$ by removing expired elements, i.e. elements that were included more than $CT$ days ago, where $CT$ is the contagious period (typically 14 days).
\end{enumerate}

\textbf{At each $encounter$ with node $B$, node $A$ (see Figure \ref{fig:proximity}):}

\begin{enumerate}
    \item collects $ EBID_{B,i} =g^{B_i}  $, from encounter $B$;

\item if the encounter satisfies some conditions\footnote{Conditions could be based on the encounter duration, number of received $EBIDs$, signal strength, etc. - to be defined with epidemiologists.}, node $A$:
\begin{enumerate}
    \item computes the duration of the encounter, $t_{A,B}$;
    \item computes 2 $PETs$: $$PET^1_{i} = H("1" ~|~g^{A_i\cdot{B_i}})$$ and $$PET^2_{i} = H("2" ~|~ g^{A_i\cdot{B_i}})$$
    \item If (the bit-string $g^{A_i} $ is greater than bit-string $g^{B_i}$)\footnote{This test can be generalized. $A$ and $B$ only need to implement a total ordering function that given 2 inputs $I_1$ and $I_2$ known by A and B outputs an order between A and B. }:
        \begin{itemize}
            \item[-] node A stores $PET^1_{i}$ in $RTL_A$ and the $(PET^2_{i}, t_{A,B}, day)$ tuple\footnote{Note that this tuple could be extended dynamically, upon request from the server, with information, such as reception signal strength, that could be useful to the health authority compute the risk score.}  in $ETL_A$\footnote{Node $B$ does the inverse, i.e. it stores  $PET^2_{i}$  in $RTL_B$ and $(PET^1_{A,i}, t_{A,B}, day)$ tuple in $RTL_B$.},
        \end{itemize}
        otherwise:
        \begin{itemize}
            \item[-] node $A$ stores  $PET^2_{i}$  in $RTL_A$ and $(PET^1_{i}, t_{A,B}, day)$ tuple in $ETL_A$\footnote{Node $B$ stores  $PET^1_{i}$  in $RTL_B$ and $(PET^2_{A,i}, t_{A,B}, day)$ tuple in $RTL_B$).}.
        \end{itemize}

\end{enumerate}
    \item deletes $g^{B_i} $;
\end{enumerate}

\subsubsection{Why using two lists, $RTL$ and $ETL$?}

Device $A$ encodes each encounter with $B$ into 2 $PETs$, $PET^1$ and $PET^2$, that are stored in two different lists, $ETL_A$ and $RTL_A$. As we will see later, one of the PET tokens is used to query the server for exposure and the other is uploaded to the server if $A$ is diagnosed-positive. 

Using two types of unlinkable $PET$ tokens for the same encouter prevents the server from linking the tokens used by $A$ in its $ESR\_REQ$ requests with the tokens that it uploads to the server (if diagnosed positive). Without this protection, the server could used these links to reconnect together the tokens of its $EList$ that belong to $A$ and derive $A$'s proximity graph.

Note also that $B$ generates the same 2 $PET$ tokens but stores them in its lists in the reverse order (see Figure \ref{fig:proximity}). Therefore the server cannot link the $PET$ token that $A$ and $B$ use in their $ESR\_REQ$ queries for their encounter, but can still perform the $PET$ matching if one of the users is diagnosed-positive and uploads her $ETL$ list.


\subsubsection{PET Generation Synchronization}

A device $A$, broadcasting $EBID_A$, that is in contact with another device $B$, broadcasting $EBID_B$, since time $t_b$ generates a new $PET$ token if:
\begin{enumerate}
 \item $A$ starts using a new $EBID$, $EBID_{A,i+1}$, at time $t_e$, or  
\item $A$ does not receive $EBID_B$ anymore since time $t_f$ for more than a given period of time (value to be defined). This scenario can occur when $B$ has changed its $EBID$, has moved away from $A$ or is temporarily behind an obstacle.
\end{enumerate}

In this case, device $A$ generates a new $PET$ to which it associates the duration\footnote{To avoid generating and storing $PET$ tokens that are very brief, it might be reasonable to only generate a $PET$ token if its encounter duration is larger than a certain value (for example, 2-3 minutes).} $(t_e-t_b)$ or $(t_f-t_b)$. According to various situations, devices may generate several $PET$ tokens for a single continuous proximity encounter. 

\begin{figure}[ht!]
\includegraphics[width=1.0\linewidth]{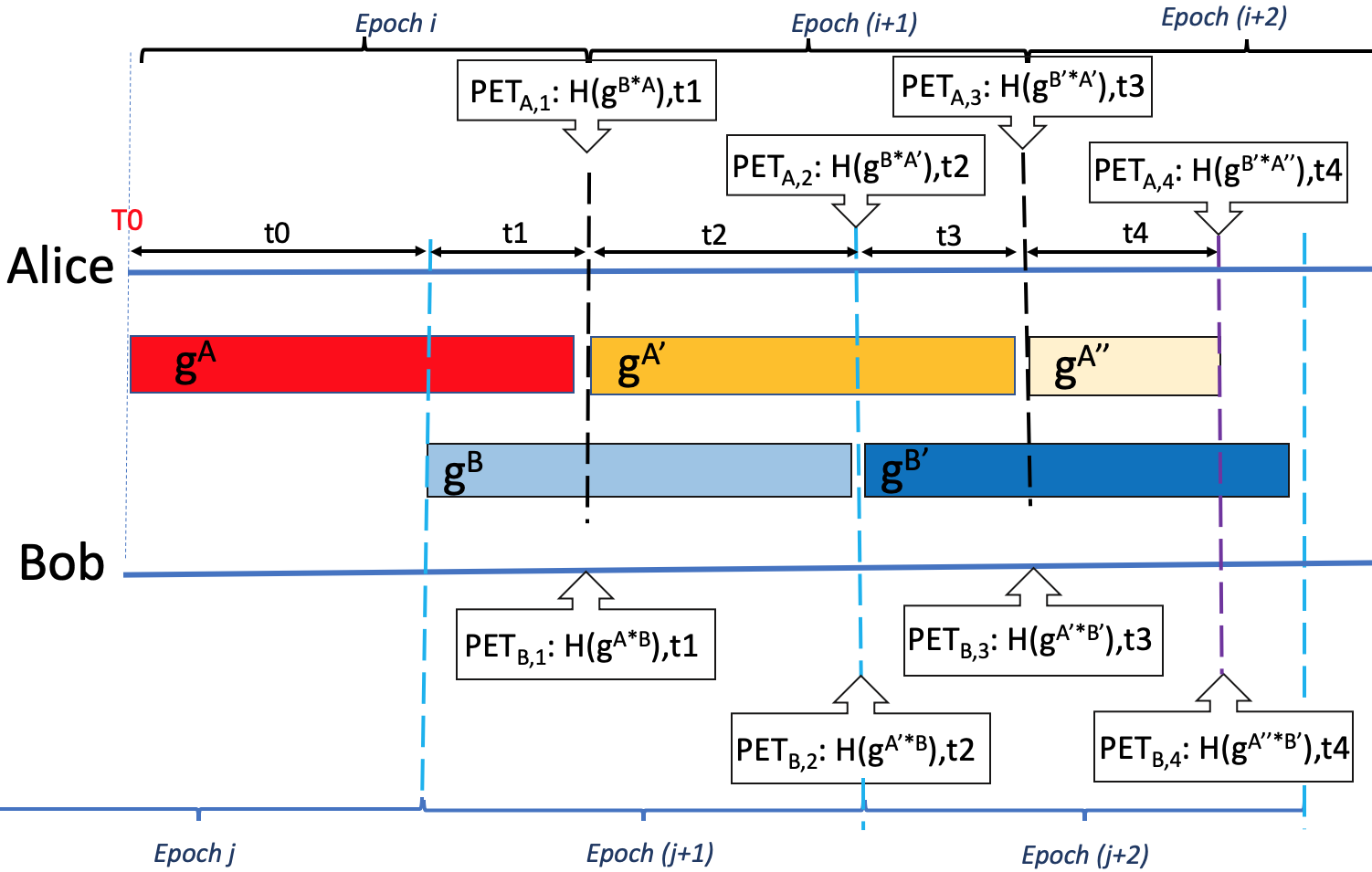}  
\caption{$PET$ Generation: this figure illustrates  $PET$ generation in $DESIRE$. The node $A$ (resp. B) generates a $PET$ token when it changes its $EBID$ or when it does not receive the $EBID$ of its encounter, node $B$ (resp. node $A$), for a giving period of time (resulting from a change of $EBID$ of node $B$ or node $B$ moving away). In this example, node $A$ starts broadcasting $g^A$ at time $T_0$. It encounters node $B$ at time $T_0+t_0$. At time $T_0+t_0+t_1$, node $A$ changes its $EBID$. As a result, node $A$ generates the $PET$ token $H(g^{B.A})$ and stores it together with its duration $t_1$. Similarly, at time $T_0+t_0+t_1$, node $B$ does not receive $g^A$ anymore: it therefore generates the $PET$ token $H(g^{A.B})$ and stores it together with its duration $t_1$, and so on.}
\label{fig:sync}
\end{figure}

 Figure \ref{fig:sync} shows various $PET$ tokens that can be generated by two devices $A$ (Alice) and $B$ (Bob), taking into account the above rules, and the loose time synchronization between devices $A$ and $B$. It allows to understand the extent to which our proposal allows anonymous proximity encounters detection in a truly decentralized context. Note that each node stores the duration of the encounter, together with the corresponding $PET$ token.

%
\subsection{Infected node declaration}
\label{sec:infected-node-decl}
\begin{enumerate}

\item

If user $U_A$ is tested and diagnosed \mbox{COVID-positive} at a hospital or medical office, she is proposed to upload each element of her $ETL_A$ list to the server (this document does not specify the interactions between $App_A$ and the health authority since it can differ from one medical system to another\footnote{One possible solution is that the user obtains an authorization code from the hospital or the medical office when it is diagnosed COVID-positive. The User can then use this code to obtain $N$ anonymous authorization tokens, where $N$ is the number of elements in $U_A$'s $ETL$ list (see Appendix \ref{sec:authorizationtoken}). $U_A$ can then used these tokens to upload each of $ETL_A$'s element one by one.}). Note that this upload is anonymous, in particular it does not reveal the identity of $U_A$ nor any of its $EBIDs$ to the server.

If insufficient precautions are taken, $ETL_A$ could potentially be used by the server to build the de-identified social/proximity graph of the infected user.  The aggregation of many such social/proximity graphs may lead, under some conditions, to the  de-anonymization of its nodes, which results in the social graphs of the users. It is therefore necessary to "break" the link between any two elements if $ETL_A$. Therefore, instead of uploading $ETL_A$, our scheme uploads each of its elements independently and in random order.
 Different solutions can be envisioned to achieve this goal:
 
\begin{itemize}
\item 
     Each element of $ETL_A$ are sent to the server one by one using a $Mixnet$\footnote{Since all mobile telecom operators are using NAT, it should be studied when the use of a Mixnet or proxy is really needed.}. Upon reception of these messages, the server won't be able to associate them with a specific $ETL$ if the upload is spread over a long period of time.
\item
    The $ETL_A$ is uploaded on a trusted server (for example at a hospital or health organization) that mixes the elements of all infected users' $ETLs$. The back-end server has only access to the exposed entries via a specific API provided by the trusted server.
\item 
    The back-end server is equipped with some secure hardware component that processes the uploads of the $ETLs$. The back-end server has only access to the exposed entries via a specific API provided by the secure hardware module.
\end{itemize}
 
    
    \item The server maintains a global list, $EList$, of all exposed $(token,day, t)$ tuples (coming from all infected users).
\end{enumerate}

%
\subsection{Exposure Status Request }


In order to check whether user $U_A$ is "at risk", i.e., if she has encountered infected and contagious users in the last $CT$ days, application $App_A$ regularly\footnote{The queries are sent regularly and at most every $ESR\_min$ epochs. If a user is allowed to perform $N$ queries per day, $ESR\_min$ is defined as $T= 86400/(N * epoch\_duration\_sec )$.} sends "Exposure Status" Requests ($ESR\_REQ$) to the server  for $ID_A$. The server then computes a "risk score" value. The server replies with a $ESR\_REP$ message that is  set to "1" when the user is "at risk" (i.e., if the "risk score" is larger than a threshold value) or to "0" otherwise.

\begin{enumerate}

\item Node $A$  periodically sends to the server, via a TLS proxy, an $ESR\_REQ_{A,i}$ message that contains $ID_A$, $EK_A$, the $PET$ tokens of $RTL_A$.

\item The server retrieves $IDTable[ID_A]$, decrypts each of its elements with $EK_A$. 

\item The server verifies if $(i-SRE_A)$, where $i$ is the current epoch number, is smaller than a threshold $ESR\_min$ (this test is meant to limit the number  of daily requests performed by $U_A$). If this is the case, the server returns an $ESR\_REP_{A,i}$ message with an error code. The server then encrypts each element of $IDTable[ID_A]$ with $EK_A$ and erases $EK_A$. Otherwise, it continues.

\item The server verifies the $UN_A$ flag. If $UN_A = true$, the server returns the same $ESR\_REP_{A,i}$ message set to "1" (at risk of exposure)\footnote{
                Note that it is good practice for an application whose user is already notified "at risk" to keep on sending $ESR\_REQ$ queries and receive $ESR\_REQ$ messages to make this application traffic indistinguishable to any other application when observing the generated network traffic.}.
The server then encrypts each element of $IDTable[ID_A]$ with $EK_A$ and erases $EK_A$. Otherwise, it continues.

\item The server then checks whether any of the $PET$ tokens of $RTL_A$ appear in any   $(token,day, t)$ tuples of $EList$. If yes, the server removes the matching tuples from $EList$ and adds all the $({day,t})$ pairs in the  $LEPM_A$ list of $IDTable[ID_A]$.

\item  The server computes an exposure risk score for user $U_A$, stores\footnote{Note that server could also store the $RTL$ tokens generated in the last $CT$ in $IDTable$. That would save some bandwidth (since a node would only need to send it daily token in its $ESR\_REQ$ queries). In addtion, this would improve security on the phone that would only store the daily tokens locally.} it in $IDTable[ID_A]$ (field $ERS_A$).

\item  Two situations are then possible:

    \begin{enumerate}
    \item 
        If the computed score indicates that the user is \atrisk, the server sets $UN_A$ at $"true"$. An $ESR\_REP_{A,i}$ message set to "1" (\atrisk)  is then returned to the user.
    \item
        If the computed score does not indicate any significant risk, an $ESR\_REP_{A,i}$ message set to "0" is returned to the user.
    \end{enumerate}
    \item After sending the reply message, the server encrypts each element of $IDTable[ID_A]$ with $EK_A$ and erases $EK_A$.

\item If $ESR\_REP_{A,i}$ is set to "1":
    \begin{itemize}
        \item[-] User, $U_A$, receives a notification from $App_A$ with some instructions (for example to go to the hospital to get tested, to call a specific number or go on quarantine).
    \end{itemize}


\end{enumerate}


Note that, for a given user $A$, since the $PET$ tokens used in the uploaded list, $ETL_A$, and in the requests, $RTL_A$ are different for the same encounters, the server can not link the elements and can not build any proximity graph. Furthermore, the only information that leaks out of a request is the number of elements, which could give some information about the number of encounters, of $RTL_A$. One solution is to set the number of elements in the request to a fixed value, $T$.
If the number of elements of $RTL_A$,  $N_A$ is smaller than $T$, $A$ pads its request with $(T-N_A)$ bogus tokens. If $N_A$ is larger than $T$, $A$ only uses $T$ elements of $RTL_A$ for its requests.
Other solutions could be to encode the elements of $RTL_A$ into a Bloom or Cuckoo filter.
%
\subsection{Notified node management}


When a node is notified at risk, its $UN_A$ flag is set to "true" in $IDTable$. From this point on, the server will process its $ESR\_REQ$ queries as usual but will keep replying with a $ESR\_REP$ message set to "1" regardless of the $ESR\_REQ$ queries. 

When set "at risk", the notified node has several options:
\begin{itemize}
\item She is tested and diagnosed as COVID-positive:
    \begin{itemize}
    \item 
        In that case she can upload her $ETL_A$ list as described in Section~\ref{sec:infected-node-decl}.
    \item 
        Independently she can inform the server that her identifier $ID_A$ was tested positive via a specific protocol (not specified in this document). This notification\footnote{Note that this notification must contain an anonymous Authorisation Token to prove that she was actually tested positive.} is done independently of the previous $ETL_A$ list upload and can not be linked together (the server cannot identify the $PET$ tokens that were uploaded by $ID_A$). 
    \end{itemize}
\item She is tested and diagnosed as COVID-negative:
    \begin{itemize}
        \item 
            She can inform the server that her identifier $ID_A$ was tested negative via a specific protocol (not specified in this document)\footnote{Informing the backend server about the results of the COVID tests could be very important to define and calibrate the risk score functions.}.  As a result, the $UN_A$ flag is reset to "false" on the server.
    \end{itemize}
\item She decides not to be tested or not to inform the server about the result of her test.
    In that case her "at risk" status will be reset automatically after a certain fixed period of time (3-4 days, value to be defined).
\end{itemize}

In any case, an application that has been notified "at risk" continues to send and receive $EBID$s and to compute $PET$s. This is required for instance when a user is waiting for a test result, if this latter turns out to be negative: encounters continue to be recorded and as soon as the user unlocks her status at the server, the updated exposure status can be computed without any gap in the history.\\

A user that was diagnosed positive should have the option to continue using the application as long as she is contagious. During this period, she must regularly upload her $ETL$ list to the server.\\

It is important to note that the above procedure needs to be discussed with epidemiologists and the health authority, and is therefore subject to modifications. 

\section{Risk Analysis}
\label{sec:analysis}

\begin{itemize}
\item[-] \textbf{Server Data Breaches:} The server only stores pseudonymous data. In addition, this information is minimized and only used to compute the exposure risk scores. Furthermore, each entries in $IDTable$ of a device $A$ is encrypted using a key $EK_A$ that is stored only by $A$ and provided to the server with the $ESR\_REQ$ queries. As a result, in case of a data breach of the server, all useful information will be encrypted. This risk associated with a data breach is then minimal.

\item[-] \textbf{Passive Eavesdropping/Tracking (by malicious users and authority):} Since PET tokens are unique per encounters and are computed locally, passive eavesdroppers only get the $EBIDs$, that are changing at every epoch. Furthermore, if the authority deploys some Bluetooth receivers, it will not be able, as a result of the decisional Diffie–Hellman (DDH) assumption, to relate any EBID (i.e.,  $g^{A_i} $) to any PET tokens. 
Passive tracking by the server or users is therefore not possible.

\item[-] \textbf{Active Eavesdropping/Tracking (by malicious users):} Active Eavesdropping/Tracking by users is not possible.

\item[-] \textbf{Active Eavesdropping/Tracking (by malicious authority):} If the authority is active and deploys Bluetooth devices that also broadcast their own $EBIDs$, 
containing for example $g^Z$,  the target device, $A$, will generate and store the $PET$ tokens $H("1",g^{A\cdot~Z})$ and $H("2",g^{A\cdot~Z})$. 
The server's devices can also generate the same tokens that the server could use to identify the target's $ESR\_REQ$ messages and possibly
 track some of his locations. Since the $ESR\_REQ$ queries of a node are linkable\footnote{The $ESR\_REQ$ contains the ID of the requesting user.}
with enough of these tracking devices, the server could possibly re-identify some users\footnote{Although this attack is technically possible, we should acknowledge that malicious authorities have probably more efficient ways to track users if they want to (such as cellphones, wifi, etc).}. This attack can be mitigated with the solution proposed in Section \ref{sec:slrobert}.

\item[-] \textbf{Reconstructing social interaction graphs (by malicious authority):} 
When a user A is diagnosed COVID-positive, he anonymously uploads all elements of its $ETL_A$ independently. Consequently, the server is not able to make any links, neither between the user and the uploaded $PET$ tokens nor between the uploaded $PET$ tokens.
When user $A$  queries the server for its exposure status, the server is able to link $ID_A$ to all $PET$ tokens contained in the $ESR\_REQ$ query. However, the $PET$ tokens used in the $ESR\_REQ$ queries are different from the tokens uploaded to the server if she ever gets diagnosed. Consequently, the server is not able to infer any links between exposed tokens in its $EList$ and tokens in requests.
Furthermore when two users $A$ and $B$ who exchanged $EBIDs$ and built associated $PET$ tokens, request the server with different tokens. The server is thus not able to link any tokens in different requests.

\item[-] \textbf{Infected Node Re-identification (by malicious user):} User can not identify infected contacts since this information is kept on the server and users only get an exposure risk score. However, the  “one entry” attack\footnote{In this attack, the adversary has only one entry, corresponding to $User_T$, in her local lists (this can easily be achieved by keeping the Bluetooth interface off, switching it on when the adversary is next her victim and then switching it off again). When the adversary is notified "at risk", she learns that $User_T$ was diagnosed COVID-positive.}, that is inherent to all schemes is still possible. This attack could be mitigated by:
\begin{itemize}
\item[-] requiring users to register in order to limit Sybil attacks;

\item[-] limiting the number of requests that each node can perform per day and limiting it even more when a user is notified "at risk". This counter-measure limits the scale of the attack;  

\item[-] using a probabilistic notification scheme, as presented in Appendix~\ref{sec:proba};

\item[-] sending a $ESR\_REP$ set to 1 only if the number of exposed tokens of the requesting user is strictly larger than 1;
\item[-] having the server verify that the requests contain at least N tokens before providing a reply "1". However, this counter-measure is not very strong since it does not prevent an adversary from using fake tokens with the target token.
\end{itemize}

%

\item[-] \textbf{Replay attacks}: not possible since it is assumed that the communication is symmetrical. For example, if a malicious node, Eve, replays the $EBID_C$ to $A$, $A$ will compute and store the corresponding $PET^1$ and $PET^2$. However, C won't have these values in his $RTL$ and $ETL$ tables.
\item[-] \textbf{Relay attacks}: only possible within, at most, one epoch.

\item[-] \textbf{False Alert Injection Attacks:} the pollution is an attack where a malicious node colludes with a diagnosed user to include the identifiers of some victims in his contact list. The goal of the malicious adversary is to make the app of a target victim raise false alerts. This attack requires that the colluding user and the victim interact (to compute their $PETs$). Such attacks are therefore only possible via a relay attack, i.e. only within an epoch.

\end{itemize}

%
%
\section{Towards a State-less DESIRE}
\label{sec:slrobert}

In $DESIRE$, the server stores some information about registered users. The information is minimal and is securely stored. We believe that this feature is a strength of our scheme since it allows to mitigate some attacks (by controlling the number of registered users and limiting the request frequency). It also allows the health authority to compute and update the risk score according to the evolving situation, it provides information to the health authority about user exposures that could be very valuable to optimize the risk score function,  as discussed Section~\ref{sec:riskScoreConsiderations}. 

Having said that, it is possible to transform $DESIRE$ into a state-less system where the server would be a mere "matching machine" between the $PET$ tokens that are uploaded by infected users and the tokens that are contained in the $ESR\_REQ$ queries. 

The protocol would operate as follows:
\begin{itemize}
    \item[-] \textbf{Application Initialization:} Users do not need to register to the server. They however need to obtain $CT$ different anonymous authorization tokens, $AT$, per day. Each of these authorisation tokens should only be valid for a specific day and could be obtained, in batch, during registration\footnote{A user would then obtain $CT$ tokens per day, under $CT$ different keys 
(and different every day: a token usable today for $d_i$, for $d_{i-1}$, for $d_{i-2}$, ... a token usable tomorrow for $d_{i+1}$, $d_i$, $d_{i-1}$, etc,). A specific signing key is 
used for each role of the token. This is in the same vein as done in on-line e-cash schemes 'a la Chaum' with blind signatures \cite{chaum}.}.
    \item[-] \textbf{Encounter Discovery}: same as in the regular $DESIRE$ protocol.
    \item[-] \textbf{Infected Node Declaration}: same as in the regular $DESIRE$ protocol.
    \item[-] \textbf{Exposure Status Request}: 
    users query the server with unlinkable $ESR\_REQ$ queries, where each of them contains a subset of different $RTL_A$'s elements. Each $ESR\_REQ$ query can, for example, contain the $PET$ tokens generated during the same day. In this case, on each day $d_i$, a user sends $CT$  different and unlinkable $ESR\_REQ$ queries  (one containing the tokens generated during $d_i$, one containing the tokens generated during $d_{i-1}$, …, one containing the tokens generated during $d_{i-CT}$). 

The server processes each of these $ESR\_REQ$ queries independently, by checking whether any tokens contained in the request appears in its list of exposed tokens, $EList$. It then compute the resulting exposure risk score of each requests and sends the result back to the requested user.
\end{itemize}
Note that with this extension, the server cannot compute the user's "global" exposure risk scores anymore but only unlinkable "daily" risk scores. The App obtains $CT$ different risk scores (for the different $CT$ previous days) that it needs to aggregate into a global one. This also implies that all apps have to include an aggregation function that might need to be updated regularly.

 Lastly, it is also possible to fully decentralize the risk score assessment by publishing to all Apps the exposed $PET$ tokens contained in $EList$ of the server. The resulting scheme would then be very similar to other so-called "decentralized" schemes, such as $DP3T$ \cite{DP3T}. 
However, we are not favorable to this approach since it would decrease resiliency against infected node re-identification. Furthermore, fully decentralizing the risk score computation could lead to the generation of many uncontrollable notifications. This could have a dramatic impact in the population, reduce the trust in the system, and consequently, its adoption rate (see Section~\ref{sec:riskScoreConsiderations}).

%% file: appendix.tex
%
\section{Authorization Token Generation}
\label{sec:authorizationtoken}

We assume that the server has a $RSA$ certificate with public key $(e,n)$ and private key $(d,n)$.

\begin{verbatim}
User ------> phone_number ------>  server 
\end{verbatim}

Server computes $ID=H(phone\_number)$ and checks that $ID$ does not exist.
Server sends a PIN code to the User via SMS (the SMS is used to verify 
that the user owns the phone number).
Note that the PIN code could be obtained via another mechanism, if the user 
does not have a sim card. For example, it could be delivered by doctors or 
the medical authority via, for 
example, email. The SMS is used to verify that the User owns the phone number.

The user generates two random numbers, $c$ and $R$, computes $c^e.H(R) \pmod{n}$ and sends:
\begin{verbatim}
User ------> PIN, c^e.H(R) (mod n) ------>  server 
\end{verbatim}

The server verifies PIN, computes $REP = (c^e.R)^d = c.H(R)^d \pmod{n}$ and sends:
\begin{verbatim}
User <------ REP = c. H(R)^d (mod n) <------ Client
\end{verbatim}
The user computes $\sigma = REP/c= H(R)^d \pmod{n} $ and obtains the authorization token $(R, \sigma)$.

Server deletes $phone\_number$ and the $REP$.

The user can then use his authorization token to prove that she is a 
registered user when sending $ESR\_REQ$ queries or during the registration phase.

However, although the user reveals her phone number (to verify that there is only one application registered per smartphone to limit Sybil attacks), the server is unable to link the user's phone number with the generated authorization token (it cannot link the token $(R, \sigma)$ to any phone number.)

Note that a user that uninstall the application, won't be able to register a new application at a later time. If this is problematic, a solution would be 
to authorize a limited number (i.e., 2 or 3) of user registrations per phone number.

\section{Towards Probabilistic Notifications}
\label{sec:proba}


As described in previous work \cite{canetti2020anonymous,cryptoeprint:2020:399}, all proximity-tracking schemes are vulnerable to the "one entry" attack. In this attack, the adversary has only one entry, corresponding to $User_T$, in her locall lists \footnote{This attack can easily be achieved by keeping the Bluetooth interface off, switching it on when the adversary is next her victim and then switching it off again.}. When the adversary is notified "at risk", she learns that $User_T$ was diagnosed COVID-positive.
As described in Section \ref{sec:analysis}, $ROBERT-v2$ proposes some mitigation measures. However, we consider that the only way to prevent this attack is to use {\em probabilistic} notifications in order to  introduce some "deniability". In such a scheme, the server that receives a $ESR\_REQ$ message would reply: 
     \begin{itemize}
        \item "0" (i.e., not at risk) if the User's ID is not in the list of exposed IDs.
        \item "1" if the User's ID is in the list of exposed IDs \textbf{or} if it is \textbf{randomly selected} by the server (the server selects additional users to receive a "1" reply with probability $p$).
    \end{itemize}   
    As a result, if the user receives a "1" back, she does not know whether it is because she has been exposed or whether she has been randomly selected by the server. Since the user cannot query the server anymore (as it already received a reply "1" back), she cannot send additional requests to refine his attack. We acknowledge that this attacks remains possible by $n$ colluding nodes that target one user. In this case, the $n$ colluding nodes will all get a "1" back and will find out the exposure status of their victim.  The scalability of the attack is however reduced since it now requires $n$ adversaries to target one victim.
    
    The side effect of this proposal is that it introduces some false positives, i.e., some people will be notified whereas they are not really "at risk" (at least according to the proximity risk score). Is this acceptable or not? There are several elements of answer to this question. First, we need to acknowledge that proximity tracing is not perfect, and that there will be anyway false positives or false negatives. In this context, is it really problematic to add $5\%$ or $10\%$ more false positives? Second, the answer to this question may also depend on what the application is used for. If the App is used to target users that should get tested, we believe that testing 5 or 10\% more users randomly should be quite acceptable. If the App is used to notify users to go in quarantine, false positives could be more problematic...

%% file: Bluetooth_SCAN_RSP.tex
\section{Bluetooth communications}
\label{sec:bluetooth_com}
Identifiers broadcasted over Bluetooth, EBIDs, are now larger than 16 bytes and can therefore not be carried by an advertisement packet alone. We propose two solutions to transmit this larger message over Bluetooth. 

\subsection{A \emph{scan response} based approach }
\begin{figure}
    \centering
    \includegraphics[width=11cm]{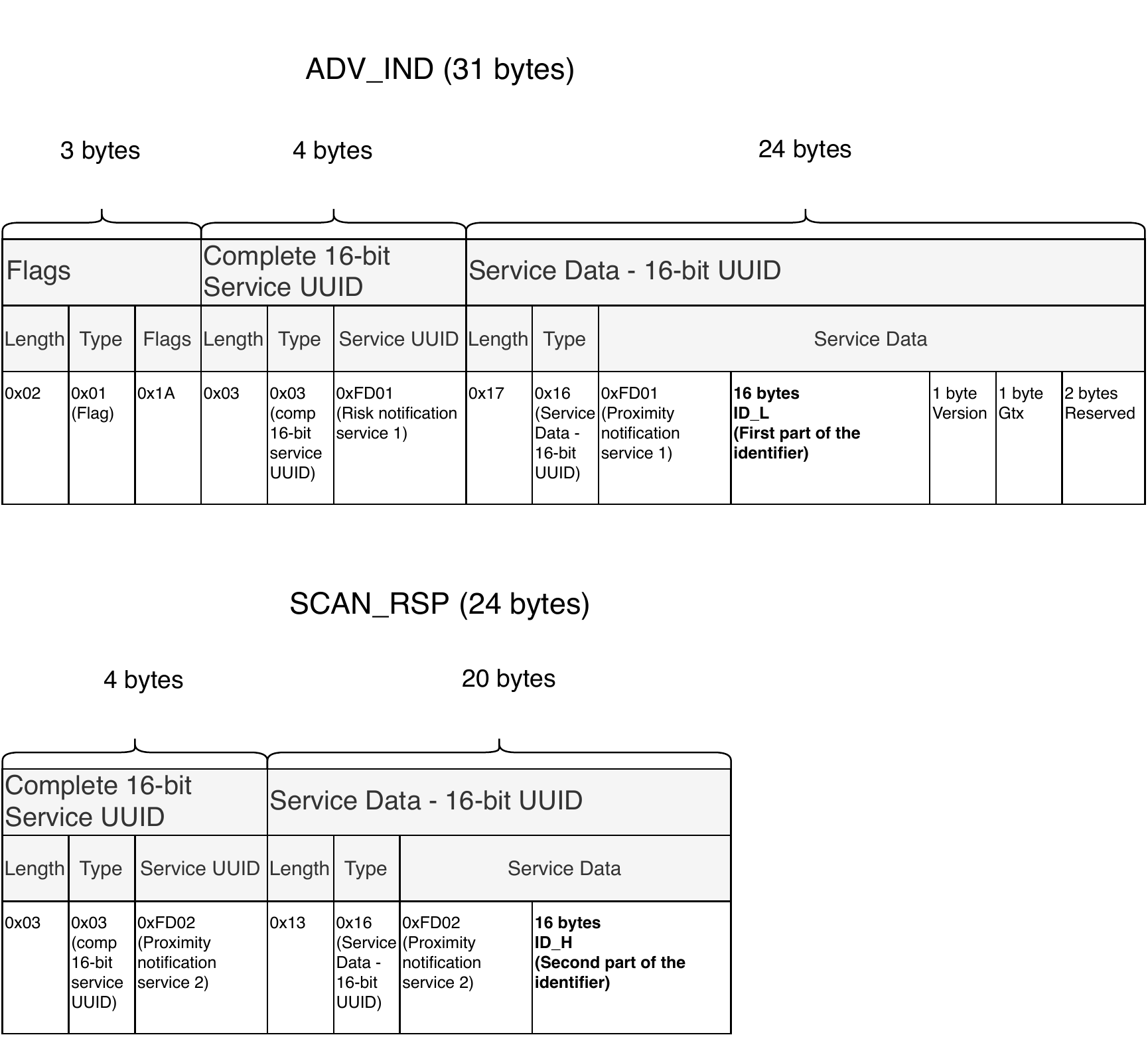}
    \caption{Payload of the \texttt{ADV\_IND} and  \texttt{SCAN\_RESP} packets used to transmit the 32 bytes identifier.}
    \label{fig:BLE_ADV_SCAN_RSP}
\end{figure}

To transmit the 256-bit identifier $\mbox{EBID}_{A,i}$ over BLE, we propose to  use the scan response mechanism of BLE~\cite[Vol 6, Part B, sec. 4.4.2.3]{bluetooth_core_specification5.1}. The 256-bit identifier is to be split into two blocks of 16-bytes, the first block is included  in the advertising data of \texttt{ADV\_IND} packets while the second block is included in the in the advertising data of \texttt{SCAN\_RSP} packets (see Figure~\ref{fig:BLE_ADV_SCAN_RSP}).

\subsubsection{Identifier segmentation}
\label{sec:identifier_segmentation}
The 256 bit identifier  $\mbox{EBID}_{A,i}$ is split in to two blocks : $\mbox{ID}_L = \mbox{LSB}_{16}(\mbox{EBID}_{A,i})$ and  $\mbox{ID}_H = \mbox{MSB}_{16}(\mbox{EBID}_{A,i})$, where $\mbox{MSB}_{16}(x)$ and $\mbox{LSB}_{16}(x)$ are functions that return respectively the 16 most significant bytes and the 16 least significant bytes of $x$.

\subsubsection{Dedicated services}
\label{sec:dedicated_services}
Each of those blocks is  configured as data for a service. To this aim, two dedicated services are defined:
\begin{itemize}
    \item  Proximity notification service 1 (PNS1), with 16bit UUID \texttt{0xFD01}\footnote{The value of those UUID is not definitive and will need to be selected in collaboration with the by Bluetooth SIG.}, will carry  $\mbox{ID}_L$ along with metadata (protocol version, corrective gain for Tx power and  reserved bytes).
    \item Proximity notification service 2 (PNS1), with 16bit UUID \texttt{0xFD02}, will carry  $\mbox{ID}_H$.
\end{itemize}

\subsubsection{Advertising and scan response payload}

The payload of advertising packets is composed of:
\begin{itemize}
    \item Flags (3 bytes)
    \item Complete 16-bit UUID (4 bytes) carrying the UUID of Proximity notification service 1 (\texttt{0xFD01}) 
    \item Service Data - 16-bit UUID (22 bytes) carrying the data for Proximity notification service 1, i.e. $\mbox{ID}_L$ (16 bytes), protocol version (1 byte), corrective gain for Tx power (1 byte), and 2 reserved bytes.
\end{itemize}

The payload of scan responses is composed of:
\begin{itemize}
    \item Complete 16-bit UUID (4 bytes) carrying the UUID of Proximity notification service 2 (\texttt{0xFD02})
    \item Service Data - 16-bit UUID (20 bytes) carrying the data for Proximity notification service 2, i.e. $\mbox{ID}_H$ (16 bytes)
\end{itemize}

We assumed that the rotation of EBID is synchronized with the device address. Therefore, the two blocks $\mbox{ID}_L$ and $\mbox{ID}_H$ can be linked via the device address. If it were not the case, an additional identifier must be included in the payload of advertising and scan response packets to allow the reconstruction of the EBID.

\subsubsection{Advertising and scanning}

As stated in Bluetooth specifications~\cite[Vol 6, Part B, sec. 4.4.2.3]{bluetooth_core_specification5.1}, after receiving an advertisement packet (\texttt{ADV\_IND} PDU), a scanner can send a scan request (\texttt{SCAN\_REQ} PDU) or request additional
information about the advertiser. If the advertiser receives a \texttt{SCAN\_REQ} PDU that contains its device address it shall reply with a \texttt{SCAN\_RSP} PDU on the same primary advertising channel index.

Devices are to be configured to follow this request response mechanism. More specifically:
\begin{itemize}
    \item A device should always send a scan request to in response to a new advertisement packet.
    \item Upon reception of a scan request, a device should always respond with as scan response.
\end{itemize}

\subsection{Fragmentation approach}
Another solution is to rely on splitting the $\mbox{EBID}_{A,i}$ in two blocks and transmitting those blocks alternatively in advertising packets. 

More specifically, the $\mbox{EBID}_{A,i}$ is  divided into two blocks $\mbox{ID}_L$ and $\mbox{ID}_H$ following the approach presented in Section~\ref{sec:identifier_segmentation}. Those blocks are  transmitted in the payload of advertisement packets (\texttt{ADV\_IND} PDU) as service data associated to a dedicated service (e.g., Risk notification service 1 with UUID \texttt{0xFD01}, as presented in Section~\ref{sec:dedicated_services}). The service data in transmitted advertising packets alternatively take the value of  $\mbox{ID}_L$ and $\mbox{ID}_H$. 


%% file: main.bbl
\begin{thebibliography}{10}

\bibitem{cryptoeprint:2020:493}
Gennaro Avitabile, Vincenzo Botta, Vincenzo Iovino, and Ivan Visconti.
\newblock Towards defeating mass surveillance and sars-cov-2: The pronto-c2
  fully decentralized automatic contact tracing system.
\newblock Cryptology ePrint Archive, Report 2020/493, 2020.
\newblock \url{https://eprint.iacr.org/2020/493}.

\bibitem{analysis}
A.~Boutet, N.~Bielova, C.~Castelluccia, M.~Cunche, C.~Lauradoux, D.~Le
  Métayer, and V.~Roca.
\newblock Proximity tracing approaches comparative impact analysis, April 2020.
\newblock
  \url{https://github.com/ROBERT-proximity-tracing/documents/blob/master/Proximity-tracing-analysis-EN-v1_0.pdf}.

\bibitem{canetti2020anonymous}
Ran Canetti, Ari Trachtenberg, and Mayank Varia.
\newblock Anonymous collocation discovery: Harnessing privacy to tame the
  coronavirus, 2020.

\bibitem{chaum}
David Chaum, Bert den Boer, Eug{\`e}ne van Heyst, Stig Mj{\o}lsnes, and Adri
  Steenbeek.
\newblock Efficient offline electronic checks.
\newblock In Jean-Jacques Quisquater and Joos Vandewalle, editors, {\em
  Advances in Cryptology --- EUROCRYPT '89}, pages 294--301, Berlin,
  Heidelberg, 1990. Springer Berlin Heidelberg.

\bibitem{Diffie1976}
Whitfield Diffie and Martin~E. Hellman.
\newblock New directions in cryptography.
\newblock {\em {IEEE} Trans. Inf. Theory}, 22(6):644--654, 1976.

\bibitem{ECDC}
{ECDC}.
\newblock {Contact tracing: Public health management of persons, including
  healthcare workers, having had contact with COVID-19 cases in the European
  Union - second update}.
\newblock Technical report, ECDC, April 2020.
\newblock
  \url{https://www.ecdc.europa.eu/en/covid-19-contact-tracing-public-health-management}.

\bibitem{Pellegrini}
{F. Pellegrini}.
\newblock {R\'eflexions sur les outils num\'eriques de suivi de contacts}.
\newblock Technical report, April 2020.
\newblock \url{https://hal.inria.fr/hal-02554672/document}.

\bibitem{NIKE}
Eduarda S.~V. Freire, Dennis Hofheinz, Eike Kiltz, and Kenneth~G. Paterson.
\newblock Non-interactive key exchange.
\newblock In Kaoru Kurosawa and Goichiro Hanaoka, editors, {\em Public-Key
  Cryptography -- PKC 2013}, pages 254--271, Berlin, Heidelberg, 2013. Springer
  Berlin Heidelberg.

\bibitem{NHSX}
{Robert Hinch et al.}
\newblock {Effective Configurations of a Digital Contact Tracing App: A report
  to NHSX}.
\newblock Technical report, NHSX, April 2020.
\newblock
  \url{https://cdn.theconversation.com/static_files/files/1009/Report_-_Effective_App_Configurations.pdf?1587531217}.

\bibitem{bluetooth_core_specification5.1}
Bluetooth SIG.
\newblock {\em Bluetooth Core Specification v5.1}.
\newblock 2019.
\newblock Accessed: 2019-08-30.

\bibitem{epione}
Ni~Trieu, Kareem Shehata, Prateek Saxena, Reza Shokri, and Dawn Song.
\newblock Epione: Lightweight contact tracing with strong privacy.
\newblock {\em CoRR}, abs/2004.13293, 2020.

\bibitem{DP3T}
Carmela Troncoso and al.
\newblock Decentralized privacy-preserving proximity tracing., 2020.
\newblock \url{ https://github.com/DP-3T/documents}.

\bibitem{troncoso:hal-01673295}
Carmela Troncoso, Marios Isaakidis, George Danezis, and Harry Halpin.
\newblock {Systematizing Decentralization and Privacy: Lessons from 15 Years of
  Research and Deployments}.
\newblock {\em {Proceedings on Privacy Enhancing Technologies}}, 2017(4):307 --
  329, October 2017.

\bibitem{encyclopedia}
Henk~C.A. van Tilborg and Sushil Jajodia, editors.
\newblock {\em Encyclopedia of Cryptography and Security}.
\newblock Springer, 2nd edition, 2011.

\bibitem{cryptoeprint:2020:399}
Serge Vaudenay.
\newblock Analysis of {DP3T}.
\newblock Cryptology ePrint Archive, Report 2020/399, 2020.
\newblock \url{https://eprint.iacr.org/2020/399}.

\bibitem{cryptoeprint:2020:531}
Serge Vaudenay.
\newblock Centralized or decentralized? the contact tracing dilemma.
\newblock Cryptology ePrint Archive, Report 2020/531, 2020.
\newblock \url{https://eprint.iacr.org/2020/531}.

\end{thebibliography}
